\documentstyle[prd,aps,floats,epsfig,times]{revtex}
\draft
\def\be{\begin{equation}}
\def\ee{\end{equation}}
\begin{document}
\vskip 4mm
\centerline{\large{\bf Periodic distribution of galaxies in generalized  
scalar tensor theory}}
\vskip 3mm
\centerline{Narayan Banerjee\footnote{narayan@juphys.ernet.in}}
\centerline{\it{Relativity and Cosmology Research Centre, Department of 
Physics, Jadavpur University, Calcutta-32, 
India}} 
\centerline{Diego Pav\'{o}n\footnote{diego@ulises.uab.es}}
\centerline{\it{Departamento de F\'{\i}sica, Facultad de Ciencias, Edificio 
Cc, Universidad Aut\'{o}noma de Bercelona,}}
\centerline{\it{E-08193 Bellaterra (Bercelona), Spain}}
\centerline{Somasri Sen\footnote{somasri@cosmo.fis.fc.ul.pt}}
\centerline{\it{ Centro de Astronomia e Astrofisica da Universidade de Lisboa (CAAUL), Departamento de F\'\i sica da FCUL, Campo Grande, 1749-016 Lisboa, Portugal.}}
\date{{\today}}
\vskip 2mm
\begin{abstract}
With the help of Nordtvedt's scalar tensor theory an exact analytic model of 
a non--minimally coupled scalar field cosmology in which the gravitational
coupling $G$ and the Hubble factor $H$ oscillate during the radiation era 
is presented. A key feature is that the oscillations are confined to the 
early stages of the radiation dominated era with $G$ approaching its present 
constant value while $H$ becoming a monotonically decreasing function of 
time. The Brans Dicke parameter $\omega$ is chosen to be a function of 
Brans Dicke scalar field so that no conflict with observational 
constraints regarding its present value arises. 
\end{abstract}
\vskip 2mm
\section{Introduction}
One of the most striking observations in recent times having a cosmological 
implication has been that of a periodicity in the distribution of galaxies. 
This possibility was pointed out by Tifft \cite{Tif} in 1977 and was 
subsequently confirmed by the deep pencil beam survey directed at the 
galactic pole as reported by Broadhurst, Ellis, Koo and Szalay (BEKS) in 
1990 \cite{beks}. The successive peaks of the galaxy distribution has been 
found to have a periodic separation of about  $128 {\rm h}^{-1}$ Mpc 
observed over a scale of $2000 {\rm h}^{-1}$ Mpc (here ${\rm h}$ is the
present value of the Hubble constant in units of 
$100$ km s$^{-1}$ Mpc$^{-1}$).\\
A spatial periodicity of this nature is indeed a bewildering and 
uncomfortable feature because in the standard cosmological models a 
spatial homogeneity (the cosmological principle) is built in. Naturally 
there have been attempts to account for this discovery in terms of an 
illusion introduced by a temporal periodicity in cosmological quantities 
like the Hubble's parameter H or the Newton's constant G, rather than a real 
spatial inhomogeneity. The first serious attempts towards this came 
immediately following the BEKS results through the work of Morikawa 
\cite{mori1,mori2} and that of Hill, Steinhardt and Turner \cite{hst}.  
Morikawa introduced a scalar field, having a tiny mass scale of order  
$10^{-31}$ eV \cite{mori2}, non--minimally coupled to gravity and showed 
that the required periodicity could possibly be obtained from this model. 
This ansatz has its problems. For example, the predicted value of $q_{0}$, 
the present deceleration parameter, from this ansatz, is positive and could 
be as high as 60. This squarely contradicts the recent 
observation of a presently accelerating universe \cite{perl}. 
Hill {\it et al.} \cite{hst} discussed the possibilities of explaining the 
BEKS results via different scenarios. An oscillating dark matter field 
or an oscillating Rydberg ``constant" were shown to be inconsistent but an 
oscillating G or an oscillating galactic luminosities could potentially 
solve the problem. Busarello {\it et al.} \cite{bus} showed that a 
non--minimally coupled scalar field can indeed produce an oscillation 
consistent with the BEKS data while a minimally coupled scalar field 
cannot.\\
  Salgado {\it et al.} analysed an oscillating $G$ model, induced 
by the oscillations of a non--minimally coupled scalar field $\phi$, which 
also has a scalar potential $V(\phi)$  \cite{ssq1}. With a reasonable choice 
of initial conditions, they carefully fixed the parameters in the model and 
numerically investigated the possible oscillations in $G$, Hubble parameter 
$H$, energy density fractions and other quantities for spatially flat and
open Friedmann--Robertson--Walker (FRW) models. The parameters were also 
fine tuned to the requirements of  primeval nucleosynthesis. They showed 
in addition \cite{ssq2} that this scalar field can also account for the 
cosmological dark matter and up to $98\%$ of the energy density of the 
Universe can be stored in the scalar field. In all these investigations 
the cosmological parameters like $H$ or $G$ oscillate at the present epoch. 
For example, in \cite{ssq1} these parameters are monotonic functions of time 
during the early stages of cosmic evolution and enters the oscillatory phase 
only in the later stages. \\
 It is worthy of mention that the present rate of variation of $G$ has a 
stringent upper bound imposed by the Viking radar echo experiment 
\cite{reas}. It was shown by Crittenden and Steinhardt \cite{critt} that 
there is an apparent conflict between these bounds and the value that is 
needed for  periodicity in the galaxy distribution. In fact they argued 
that the nucleosynthesis constraints, consistent with these bounds, are 
very stringent on this oscillating $G$ model unless a ``fine tuning" of 
oscillation of the scalar field is assumed. In a recent work, Gonz\'{a}lez 
{\it et al.} \cite{ssq4} showed that with this restriction on the upper 
bound on the variation of $G$, one cannot have sufficient oscillation in 
order to explain the galactic periodicity as observed by the deep pencil 
beam survey.\\
Furthermore, the luminosity--redshift relation of distant supernovae led 
to the conclusion that the present universe is accelerating its expansion
\cite{perl} although, in order to facilitate the primeval nucleosynthesis, 
it must have been decelerating in the radiation era. These results might 
lead to the possibility that if the observed periodicity in the distribution 
of galaxies has to be attributed to the imprints of some temporal 
oscillations in some cosmological parameters, these oscillations should 
have taken place in the early stages of the evolution rather than recently.\\
 Keeping this possibility in mind, we present an exact analytic model of a 
non--minimally coupled scalar field cosmology in which the oscillatory 
behaviour takes place in the past, namely during the radiation era. 
We start from Nordtvedt's generalization of Brans Dicke (BD) theory 
\cite{brans} where the dimensionless BD parameter $\omega$ is taken to be a 
function of the scalar field $\phi$ \cite{berg}. We do not employ any 
additional scalar potential $V(\phi)$. The effective self--interaction is 
taken care of by the functional dependence of $\omega$. With the matter 
distribution taken in the form of radiation, i.e., $\rho = 3 p $, where $p$ 
and $\rho$ are the pressure and energy density of the cosmic fluid, it is 
found that there exists an exact analytical set of solutions for the field 
equations where $\phi$, $G$, and $H$ have an oscillatory phase. One 
important feature is that these oscillations die out in the radiation 
dominated period itself with $G$ approaching a constant value and 
$H$ becoming a monotonically decreasing function of time. The functional 
dependence of $\omega$ on $\phi$ may be carefully chosen so as to meet the 
observational requirements like the gravitational constant $G$ stabilizing 
at the present value resulting in the transition to general relativity. 
This avoids the  conflict between the bounds on variation of $G$ and 
oscillations in $G$ required for galactic periodicity. \\
The gravitational field equations in the generalized scalar tensor theory 
are too involved. So we adopt the following strategy. We effect a conformal 
transformation for the metric tensor components which results in a major 
simplification of the field equations \cite{dicke} as they become more 
tractable. This transformed version is not ``physical" in the sense that 
the geodesic equations are not valid in this version (Einstein's frame). 
For a modern review, see Faraoni {\it et al.} \cite{farao}). But as we have 
the complete analytic solutions for the equations, we can transform the 
metric back to its original version (BD frame). We discuss all the 
cosmologically relevant functions in this physical version, and so the 
results obtained can be stated with confidence.\\
In section 2 we present the model and obtain the solutions. In section 3 the 
oscillatory behaviour is studied in detail. Finally, section 4 summarizes 
our conclusions and suggests directions of further investigation.

\section{Field equations and formulation of the model}

We start from the action 
\be
\bar{\cal{S}}=\frac{1}{16\pi G_0}\int\sqrt{-\bar{g}} \left[\phi\bar{\cal{R}}-
\frac{\omega (\phi)}{\phi}\phi_{,\alpha}\phi^{,\alpha}
+\bar{\cal{L}}_{m} \right]d^4 x \ ,
\ee
where ${\bar{\cal{L}}}_m$ is the matter lagrangian and $G_{0}$ the Newtonian
gravitational constant. The dimensionless Brans-- Dicke parameter $\omega$ 
is now assumed to be a function of the scalar field $\phi$. This 
generalization of BD theory, proposed by Nordtvedt \cite{berg}, includes a 
host of non--minimally coupled scalar tensor theories, suggested from 
different physical motivations and they are in fact the special cases of 
this ansatz.\\
To simplify the calculations we effect the conformal transformation
\be
g_{\mu\nu}=\phi\bar{g}_{\mu\nu},
\ee
so that the action looks like
\be
{\cal S}=\frac{1}{16\pi G_0}\int\sqrt{-g} \left[{\cal R}-\frac{2\omega+3}
{\phi^2}\phi_{,\alpha}\phi^{,\alpha}+{\cal L}_{m} \right]d^4 x \ .
\ee
Variables with and without an overhead bar are in the original version and 
the conformally transformed version respectively. For an FRW spacetime the 
gravitational field equations in the latter version are
\be
3\frac{{\dot a}^2}{a^2}+3\frac{k}{a^2}=8\pi G_0\rho+\frac{2\omega+3}{4}
{\dot\psi}^2,
\ee
\be
2\frac{\ddot a}{a}+\frac{k+{\dot a}^2}{a^2}=-8\pi G_0 p-\frac{2\omega+3}{4}
{\dot\psi}^2,
\ee
where $a$ is the scale factor, $k$ the curvature index having values $0$ 
or $\pm 1$ and $\psi=\ln(\frac{\phi}{\phi_0})$, $\phi_0$ being a constant. 
The density and pressure of the cosmological perfect fluid are $\rho$ 
and $p$, respectively. The wave equation for the scalar field is given by
\be
(2\omega+3) ({\ddot \psi}+3\frac{\dot a}{a}{\dot \psi})=8\pi G_{0} T-
\dot{\omega} {\dot \psi}.
\ee
Here $T = \rho - 3p$ is the trace of the energy momentum tensor of 
the matter field.\\
The position of the first acoustic peak in the anisotropy power spectrum 
of the cosmic microwave background radiation strongly suggests that the 
Universe is spatially flat (see e.g., \cite{CMB}), thereby we shall 
assume $k=0$. Also  we require to investigate the behaviour of the model 
at an early stage of evolution rather than at the present epoch, so we 
take the perfect fluid distribution as radiation, i.e., an equation of 
state $\rho = 3 p$. This leads to a straightforward first integral of the 
wave equation as
\be
(2\omega+3)^{\frac{1}{2}}a^3{\dot \psi}=A_1
\ee
$A_1$ being a constant of integration. A combination of equation (4), (5) 
and (7) yields 
\be
\frac{\ddot a}{a}+\frac{\dot a^2}{a^2}+\frac{A^2}{a^6}=0,
\ee
where $A^{2} = A_{1}^{2}/12$. After integrating once, equation (8) can be 
written as 
\be
u^2=\frac{A^2}{a^4}+\frac{B_1}{a^2},
\ee
where $u(a)=\dot a$ and $B_1$ is a constant of integration. This equation 
can again be integrated and it yields three results, depending on the 
signature of $B_1$. In what follows we shall take $B_1=B^2>0$. In this case 
the solutions for the scale factor can be written in terms of the 
transcendental function
\be
a=\frac{1}{\sqrt{m^2+a^2}}\left[2B(t+\tau)+m^2\ln(a+\sqrt{m^2+a^2})\right],
\ee
where $m = A/B $ and $\tau$ is a constant of integration. From equation 
(7) and (9) one can easily obtain
\be
\left({\frac{2\omega+3}{12}}\right)^{1/2} \frac{d\psi}{da}=
\frac{m}{a\sqrt{m^2+a^2}},
\ee
>From this equation $\psi$ (or $\phi=e^\psi$) can be expressed as a function 
of $a$, if a definite functional form of $\omega=\omega(\phi)$ is chosen. 
In what follows, we take
\be
2\omega+3=\frac{3}{4C}\frac{\phi^2}{(1-\phi)[C(\phi-1)+1]},
\ee
where $C$ is a constant. \\
It deserves mention that Quevedo {\it et al.} \cite{ssq3} have shown that in 
order to account for the nonbaryonic dark matter as a non--minimally coupled 
scalar field, the effective $\omega$ parameter should be a ratio of 
quadratics in $\phi$. The action integral (1) suggests that the gravitational 
coupling $G$ is given as
\be
G=\frac{G_{0}}{\phi}.
\ee
It is well--known that $\omega\rightarrow\infty$ is a necessary (although 
not sufficient \cite{sn}) requirement for the scalar tensor theories 
becoming indistinguishable from general relativity, where $G=G_{0}$. Our 
choice of $\omega$ indicates that for $\phi\rightarrow 1$, $\omega$ goes to 
the desired infinity limit and $G\rightarrow G_{0}$ (see Eq. (12)). Using 
the expression for $\omega$ in equation (12) and the relation 
$\psi=\ln(\phi/\phi_{0})$, equation (11) can be integrated to yield
\be
\phi=1-\frac{1}{C}\sin^2\left[2~C\ln\frac{a}{m+\sqrt{m^2+a^2}}\right],
\ee   
where a constant of integration is put equal to nought. It is clear from 
equation (14) that $\phi$ is not a monotonic function of the scale factor, 
but rather has an oscillation. But in order to understand the behaviour of 
the model, one has to find the expression of the scale factor in the 
original version of the theory, i.e., $\bar a$ as given in equation (1), in 
place of $a$, because only in that version the theory retains the principle 
of equivalence and quantities carry their usual physical significance. 
Although we have an analytic form for $a$, but it is expressed in a 
transcendental way (equation (10)) and so we shall endeavour to investigate 
the actual behaviour of different relevant quantities by graphical 
representation.

\section{Oscillatory behaviour of ${\phi}$, $G$ and $H$}

\noindent It is easy to recast the expression for the scalar field $\phi$ 
in equation (14) in terms of the original scale factor $\bar a$ with the 
help of equation (2) as 
\be
\phi=1-\frac{1}{C}\sin^2\left[2~C\ln\left[\frac{\bar{a}\sqrt{\phi}}
{1+\sqrt{1+\phi\bar{a}^2}}\right]\right],
\ee
where we have chosen $B$ to be $\frac{1}{2}$ and $m$ to be $1$ for the sake 
of computational simplicity. The expression for the effective 
gravitational constant $G(=\frac{1}{\phi})$, if the present value $(G_0)$ is 
taken to be unity, is given as
\be
G=\frac{1}{\phi}=\frac{1}{1-\frac{1}{C}\sin^2\left[2~C\ln\left[\frac
{\bar{a}\sqrt{\phi}}{1+\sqrt{1+\phi\bar{a}^2}}\right]\right]}. 
\ee
The Hubble parameter $\bar H$ in this frame is given by,
\begin{eqnarray}
\bar H&=&\frac{\dot{\bar a}}{\bar a}=\frac{d}{dt}[\ln(a\sqrt{\phi})]
=\frac{\dot a}{a}-\frac{1}{2}\frac{\dot \phi}{\phi}
\nonumber\\
&=&\frac{1}{2\phi^{\frac{3}{2}}{\bar a}^3}\left[\sqrt{1+\phi{\bar a}^2}
+\frac{\sin 2X}{1-\frac{1}{C}\sin^2X}\right] \, ,
\end{eqnarray}
where $X=2~C\ln\frac{\bar{a}\sqrt{\phi}}{1+\sqrt{1+\phi\bar{a}^2}}$.
Now the scalar field $\phi$, the Newtonian constant $G$ and the Hubble 
parameter $\bar H$ are all, in principle, known in terms of scale factor 
$\bar a$. But as these relations are involved and it is hardly possible to 
investigate the nature right from these expressions, we plot these quantities 
$\phi, G$ and $\bar H$ against $\bar a$.\\

Before we actually plot these variables, the one arbitrary constant in 
equation (11), namely $C$, should be evaluated. As we are interested in the 
behaviour of the model in the early stages of the evolution, i.e., when the 
scale factor has small values, it would be useful to have a series expansion 
of equation (10). Along with the choice $B=\frac{1}{2}$ and $m=1$, equation 
(10) can be approximated to 
\be
a= \left[\textstyle{3\over{2}}(t+\tau) \right]^{1/3},
\ee
where the series is retained up to third order in $a$. Now with the help 
of  this expression, the time evolution of the scalar field can be 
expressed as follows 
\be
\phi=1-\frac{1}{C}\sin^{2} \left[\frac{2C}{3}\ln(1+\frac{t}{\tau}) \right].
\ee
Similarly from equations (17), (18) and (19) the time variation for the 
Hubble parameter $\bar H$ is
\be
\bar H=\frac{1}{3\tau(1+\frac{t}{\tau})}\left[1+\frac{1}{1-(\frac{2t}
{3\tau})^2}\sin~2\left(\frac{2C}{3\tau}t\right)\right].
\ee
Clearly the part within the square bracket produces an oscillatory behaviour 
in $\bar H$. The first part $\frac{1}{3\tau(1+\frac{t}{\tau})}$ primarily 
behaves as $t^{-1}$, which we expect to get towards the end of the 
radiation regime. We designate the functional dependence of the 
Hubble parameter at the end of radiation era as
\be
\bar H_r=\frac{1}{3\tau(1+\frac{t}{\tau})}.
\ee
Keeping in mind the relation $1+z=a^{-1}$, it is straight forward to relate 
$z_r$, the usual redshift of the expanding universe in the radiation era, to 
the redshift $z$, corrected for oscillation\cite{hst,bus}, by
\be
\frac{d{\bar z}}{d{\bar z_r}}\sim\frac{d{\bar z}}{\bar z_r}=
\frac{\bar H}{\bar H_r}= 1+\frac{1}{[1-
(\frac{2t}{3\tau})^2]}\sin~2\left[\frac{2C}{3\tau}t\right].
\ee\\

From equation (22) it is evident that the frequency of oscillation is given 
by 
\be
\nu=\frac{2~C}{3~\tau}.
\ee

\begin{figure}[hb]
\centering
\leavevmode\epsfysize=6.5cm \epsfbox{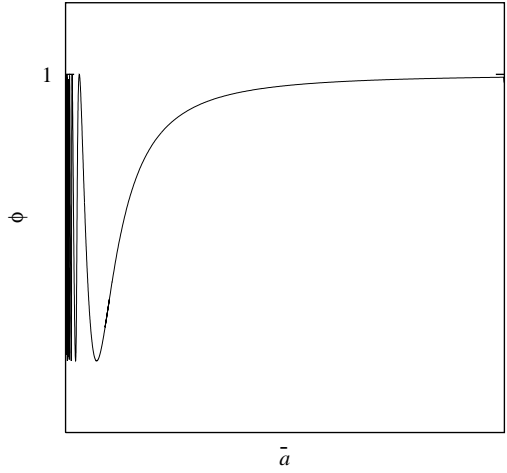}
\caption{Plot of the Brans Dicke field $\phi$ against the original scale 
factor $\bar a$}
\label{fig1}
\end{figure}
\begin{figure}[hb]
\centering
\leavevmode\epsfysize=6.5cm \epsfbox{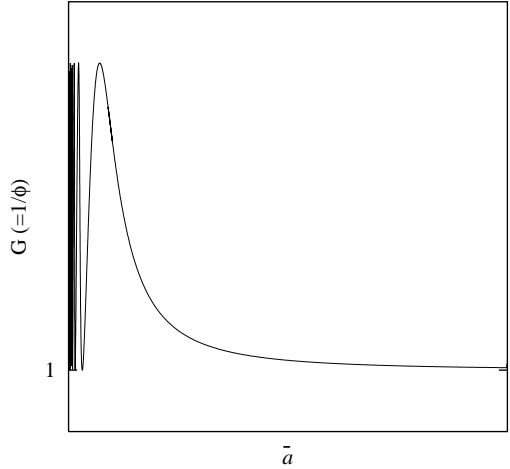}
\caption{Plot of the Gravitational Constant $G$ against the original scale 
factor $a$}
\label{fig2}
\end{figure} 

where as the amplitude of oscillation depends on time as 
\be
{\cal A}=\frac{1}{1-\frac{4t^2}{9\tau^2}}.
\ee  
We now plot $\phi, G$ and $\bar H$, all against $\bar a$ from equations (15), 
(16) and (17) respectively with a value {\bf $C=5$}.\\

Figure 1 shows that $\phi$ has an oscillation for very small $\bar a$ and 
then attains a value  of unity. From figure 2 one can find that $G$ also has 
an initial oscillation and then quickly settles down to a value close to 
its present value ($G_{0}$ is taken to be unity in the units in which the 
equations are written). $\bar H$ too has an oscillation (see figure 3), 
about a steadily decreasing mean value and then becomes a monotonically 
decreasing function of $\bar a$. This is a desirable feature as in the 
radiation paradigm we need a decelerating universe ($q>0$). Figure 4 shows 
that $\Omega_{\phi}$, the dimensionless density parameter corresponding to 
the scalar field, also has an initial oscillation, and then approaches the 
value zero in the early  radiation era. This is safely below the permitted 
upper bound of $\Omega_{\phi} < 0.2$ conducive for a successful 
nucleosynthesis \cite{varun}. \\

One point to note here is that the nature (some oscillation at an early 
stage and then monotonic behaviour) of these graphs are not really too 
sensitive to the value of $C$, only the frequency and amplitude of 
oscillations get modified with the choice of $C$.\\  

\begin{figure}[hb]
\centering
\leavevmode\epsfysize=7.0cm \epsfbox{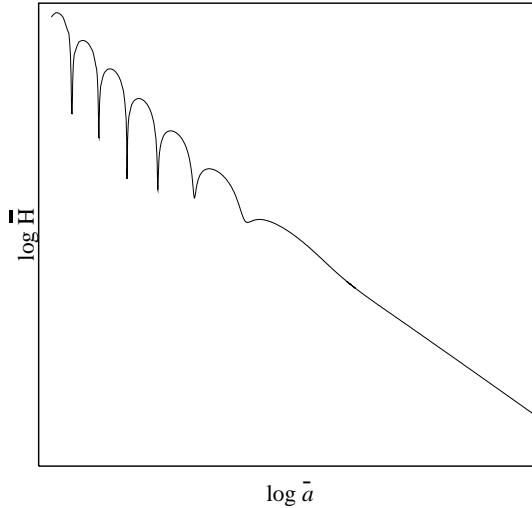}
\caption{Plot of the Hubble parameter $\bar H$ (original unit) against the 
original scale factor $\bar a$ in logarithmic scale}
\label{fig3}
\end{figure}
\begin{figure}[hb]
\centering
\leavevmode\epsfysize=6.5cm \epsfbox{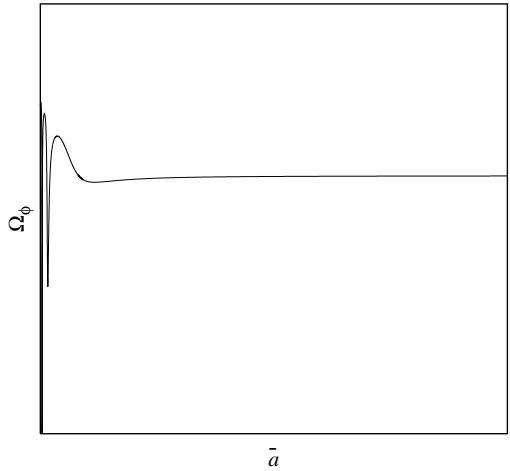}
\caption{Plot of  density parameter $\Omega_{\phi}$ against the original 
scale factor $\bar a$}
\label{fig4}
\end{figure}

\section{Concluding remarks}

In order to have a compromise between the periodicity in the galaxy 
distribution and the cosmological principle, one might require a temporal 
periodicity in the cosmological model resulting in an apparent spatial 
periodicity. An oscillating $G$ model is indeed one of the most viable 
options in this connection. \\

We have resorted to Nordtvedt's scalar tensor theory to build a model in 
which $\phi$, $G$ and $H$ present oscillations but confined to the 
radiation--dominated era of cosmic expansion. Owing to the fact that the 
dark--matter component does not interact with the radiation field it clusters 
with the periodicity of the oscillations. Later on, during the matter era the 
baryonic component is free to fall in the potential wells created by the dark 
component. Our solutions are analytical, but a bit involved -Eqs. (15)--(17). 
This is why we have depicted their behaviour with expansion. Aside from being 
compatible with primordial nucleosynthesis bounds our model has the advantage 
that the BD parameter diverges for large $\bar a$, therefore no conflict with 
local measurements of $\omega$ arises \cite{reas}. So one really has wider 
options to account for the periodicity of the galaxy distributions as an 
imprint of a temporal oscillation in $G$ and $H$.

Obviously, since the oscillations do not extend to the matter era the 
model presented in this paper cannot account for the alleged periodicity
in length of the solar year, supposedly derived from periodic structures 
found in coral fossils and marine bivalves \cite{sis}. However, it should
be noted that such periodicity may well be purely biological in origin and 
therefore unconnected to alleged oscillations of the solar year.\\

It also deserves mention that the functional dependence of $\omega$ on 
$\phi$ taken up in this work -Eq. (12)- is by no means unique. There could 
well be possibilities of $\omega = \omega(\phi)$ other than the one adopted 
here. This indeed gives a flexibility for improvisation if required to fit 
in other observations such as the late time acceleration of the Universe. \\

In fact it will be worthwhile to figure out the correct $\omega = 
\omega(\phi)$ which serves all these purpose, namely drives an oscillation 
at some stage, gives perfect ambience for nucleosynthesis, and generates 
sufficient negative pressure in the later stage so that an accelerated 
expansion for the present universe could be explained.

From the figures it can be seen that our model does not produce
enough number of oscillations, i.e., more rings of galaxies are
observed than oscillations our model is able to provide. This is why
our model cannot be viewed as fully accounting for the inhomogeneous
distribution of galaxies.  We believe, however, it may serve as a
starting point for more detailed models that overcome this
limitation. A lesser difficulty refers to the fact that our universe
seems to be accelerating its expansion today (see e.g. \cite{saul} 
and references therein) while our predicts deceleration at the present
time.  However, this may be solved by introducing some quintessence
scalar field such that its contribution to the total energy density
becomes relevant only recently \cite{quintessence}. We have chosen
not to go into that at this stage in order to focus on the problem of
galaxy distribution.

\section*{Acknowledgments} 
Thanks are due to Fernando Atrio for discussions and comments on a
earlier draft of this work as well as to Daniel Sudarsky and Marcelo
Salgado for correspondence. Likewise we  would like to thank D.
Choudhury to let us use his Fortran program for the plotting. This
work has been partially supported by the Ministry of Science and
Technology of the Spanish Government under grant BFM 2000--C--03--01
and 2000--1322.

\end{document}